\documentclass[preprint,nofootinbib,epsf]{revtex4}
\usepackage{amssymb}
\usepackage[dvips]{graphicx}
\renewcommand{\thefootnote}{\fnsymbol{footnote}}

\newcommand{\lt}   {\left}
\newcommand{\rt}   {\right}
\newcommand{\beq}{\begin{equation}}
\newcommand{\eeq}{\end{equation}}
\newcommand{\bea}  {\begin{eqnarray}}
\newcommand{\eea}  {\end{eqnarray}}
   
\newcommand{\ov}   {\overline} 
\newcommand{\la} {\langle}
\newcommand{\ra} {\rangle}

\begin{document}
\preprint{BA-06-11}
\title{Flux of Primordial Monopoles}
\vskip 8cm
\author{Shahida Dar}
\author{Qaisar Shafi}
\author{Arunansu Sil}
\affiliation{Bartol Research Institute, Department of Physics and Astronomy, 
University of Delaware, Newark, DE 19716, USA}
\begin{abstract}
We discuss how in supersymmetric models with $D$ and $F$-flat directions, 
a primordial monopole flux of order $10^{-16} - 
10^{-18} {\rm cm^{-2}}{\rm sec^{-1}}{\rm sr^{-1}}$ 
can coexist with the observed baryon asymmetry. A modified 
Affleck-Dine scenario yields the desired asymmetry if the 
monopoles are superheavy ($\sim 10^{13}-10^{18}$ GeV). 
For lighter monopoles with masses $\sim 10^{9}-10^{12}$ GeV, the baryon 
asymmetry can arise via TeV scale leptogenesis. 
\end{abstract}
\maketitle  

\renewcommand{\thefootnote}{\arabic{footnote}}
\setcounter{footnote}{0}
Unified theories of elementary particle interactions \cite{Salam:1973uk, Georgi:1974sy} 
predict the existence of topologically stable magnetic monopoles \cite{Polyakov:1974ek} 
with masses and magnetic charges that depend on the details of the underlying theory. 
In supersymmetric SU(5), for instance, the lightest monopole with mass of order 
$M_G / \alpha_G$ ($\simeq 5 \times 10^{17}$ GeV) 
carries one unit $(2 \pi/e)$ of Dirac magnetic charge \cite{Daniel:1979yz}. (Here $M_G 
\simeq 2 \times 10^{16}$ GeV denotes the gauge coupling unification scale, 
and $\alpha_G \approx 1/25$ is the unified coupling constant at $M_G$). 
The monopole also has screened color magnetic fields \cite{Daniel:1979yz}, 
and the presence of baryon and lepton number violating superheavy gauge bosons 
leads to monopole catalysis of nucleon decay \cite{Rubakov:1981rg}. 
If the monopole arises from a partial unified model such as 
$G_{422} \equiv SU(4)_c \times SU(2)_L 
\times SU(2)_R$ \cite{Salam:1973uk} (or $G_{333} \equiv 
SU(3)_c \times SU(3)_L \times SU(3)_R$), it carries two \cite{Lazarides:1980cc}
(or three \cite{Shafi:1984wk}) quanta of Dirac charge\footnote{Multiply charged monopoles also arise in 
superstring theories \cite{Wen:1985qj, Witten:2002wb}.}. 
The absence in these models of gauge bosons which mediate proton decay 
means that, in principle, the symmetry breaking scale ($M$) can be considerably below $M_{G}$ 
\cite{Leontaris:2000hh}, leading to lighter ( $\sim 10^{10}$ GeV) monopoles. In a supersymmetric 
framework intermediate scales are allowed provided proton decay via dimension five operators 
is adequately suppressed. Models with intermediate scales have 
acquired renewed interest because they naturally appear from 
compactification of superstring theories on Calabi-Yau manifolds and 
orbifolds \cite{Candelas:1985en}, from $D$-brane constructions \cite{Leontaris:2000hh, Prikas:2003ce}, 
or from intersecting brane models \cite{Aldazabal:2000cn}.
 
The fate of primordial monopoles is very closely linked to the history of 
the very early universe. In hot big bang cosmology without inflation there is 
a serious cosmological monopole problem pointed out a long time ago 
\cite{Zeldovich:1978wj}. Here we list some proposed solutions:

\begin{itemize}

\item An inflationary epoch \cite{Guth:1980zm, Linde:1981mu} reduces the primordial monopole number density 
to utterly negligible values. The subsequent transition to radiation epoch 
may give rise to thermally produced monopoles, provided the reheat 
temperature is not too far below the monopole mass \cite{Turner:1982kh}. 

\item If the symmetry breaking which gives rise to monopoles 
experiences only the last $30$ or so $e$-foldings of inflation 
\cite{Shafi:1984tt, Shafi:2006cs}, a measurable flux of monopoles comparable to
 or somewhat below the Parker bound \cite{Parker:1970xv} and the MACRO experiment limits 
\cite{Ambrosio:2002qq} may be present in our galaxy. A different scenario 
with similar objectives is discussed in \cite{Lazarides:1993fi}.

\item Monopoles and antimonopoles get linked either by electromagnetic 
\cite{Langacker:1980kd} or $Z$ flux tubes \cite{Lazarides:1980va} and 
efficient annihilation occurs as a result of rapid contraction of these tubes. Another 
resolution may be non-restoration of grand unified symmetry at 
high temperature \cite{Dvali:1995cj}. In \cite{Dvali:1997sa}, it is proposed that if 
monopoles are produced in association with domain walls, the latter can sweep away 
monopoles and then disappear. A 
black hole solution of the cosmological monopole problem is discussed in \cite{Stojkovic:2004hz}.

\item Monopoles arise in supersymmetric models with $D$ and $F$-flat directions 
\cite{Lazarides:1985bj, Lazarides:1986rt}, in which thermal 
inflation \cite{Lazarides:1985ja, Yamamoto:1985rd, Binetruy:1986ss, 
Lazarides:1987yq, Lazarides:1992gg, Lyth:1995ka} is followed by a huge release of entropy 
\cite{Lazarides:1986rt, Barreiro:1996dx}. 
An initially large monopole number density can be reduced to the Parker bound or somewhat 
below it \cite{Lazarides:1986rt}.

\end{itemize}

In this paper we wish to explore this last scenario, previously discussed 
in \cite{Lazarides:1986rt}, in which 
monopoles appear after thermal inflation is over and a huge amount of entropy is released. 
While this certainly helps with monopole dilution, 
our main challenge is to identify a framework in which an 
observable flux of primordial monopoles is compatible with the observed baryon 
asymmetry. This is particularly challenging for superheavy monopoles with 
a correspondingly large symmetry breaking scale, so that the final temperature after 
thermal inflation can be quite low, of order a few MeV. Following \cite{Stewart:1996ai}, we 
employ a modified Affleck-Dine (AD) scenario to realize the observed baryon asymmetry. 
Although our discussion is mainly focused 
on monopoles with mass  $\sim 10^{14}-10^{15}$ GeV, it can be adapted, as we will show, 
to monopoles that are somewhat heavier ($\sim 5 \times 10^{17}$ GeV) 
or significantly lighter ($\sim 10^{9}-10^{13}$ GeV) \cite{Kephart:2006zd}.

Consider the superpotential
\beq \label{spot}
W_{inf} \supset \lambda\frac{\lt(\phi\ov {\phi}\rt)^{n}}{M_*^{2n-3}} + 
\beta H_u H_d \frac{\lt(\phi\ov {\phi}\rt)^{m}}{M_*^{2m-1}}\,,
\eeq
where the scalar components of $\phi$ and ${\ov {\phi}}$, called flatons \cite{Yamamoto:1985rd}, acquire non-zero 
vevs ($M$) and break the underlying gauge symmetry $G$. $M_*$ is the cutoff scale, $H_{u,d}$ 
are the MSSM Higgs doublets, and $n$, $m$ are integers suitably chosen to yield 
the desired symmetry breaking scale $M$ and the MSSM $\mu$ term respectively \cite{Lazarides:1985bj, 
Stewart:1996ai, Jeannerot:1998qm, Dar:2003cr, Dar:2005hm}. 
The second term in Eq. (\ref{spot}) also plays an important role in 
reheating after thermal inflation.
   
The zero temperature effective potential of flatons (we use the same notation for the 
superfield and its scalar component) is given by
\beq \label{epot}
V(\phi)\,=\,\mu_{0}^4\,-\,M_s^2\,\lt(\lt|\phi\rt|^2+\lt|{\ov \phi}\rt|^2\rt)
\,+\,n^2\,\lt|\frac{\lambda \phi^n {\ov \phi}^{n-1}}{M_*^{(2n-3)}}\rt|^2
\,+\,n^2\,\lt|\frac{\lambda \phi^{n-1} {\ov \phi}^n}{M_*^{(2n-3)}}\rt|^2
\,+\,A_{\lambda} \lambda\frac{\lt(\phi\ov {\phi}\rt)^{n}}{M_*^{2n-3}}\,+\,{\rm c. c}\,,
\eeq 
where $\mu_{0}^4$ is the false vacuum energy density, such that 
$V(M)=0$ at $|\la \phi \ra|=M$, $M_s$ is the soft supersymmetry breaking mass parameter, and 
$\lt|A_{\lambda}\rt| \lesssim M_s$. Minimizing the effective potential along 
the $D$-flat direction $\lt|\la \phi \ra\rt|= \lt| \la {\ov \phi} \ra \rt|$ 
yields the symmetry breaking scale\footnote{Here the phases $\epsilon$, $\alpha$ 
and ${\ov \alpha}$ of $A_{\lambda}$, $\phi$ and ${\ov \phi}$ are taken to satisfy the relation 
$\epsilon+n\alpha+n {\ov \alpha}=\pi$.},
\bea \label{int}
M =\lt|\la\phi\ra\rt| &=&\lt[\frac{M_*^{(2n-3)}}{2(2n-1)n\lambda} \lt\{\lt|A_{\lambda}\rt|+
 \sqrt{\lt|A_{\lambda}\rt|^2\,+\,4(2n-1) M_s^2}\rt\}\rt]^{1/2(n-1)}\,,\nonumber \\
&\sim& \lt[\frac{M_s M_*^{(2n-3)}}{\sqrt{(2n-1)}n \lambda}\rt]^{1/2(n-1)}\,,
\eea
where $n \ge 2$ and for simplicity, we assume $\lt|A_{\lambda}\rt| < M_s$. Some 
typical values of $M_s$, $M_*$ and $M$ that we will consider in this paper 
are listed in Table I. 
For non-zero temperature $T$ the effective potential gets an additional contribution given by 
\cite{Dolan:1973qd} 
\beq \label{tpot}
V_{T}(\phi)=\left(\frac{T^{4}}{2\pi^{2}}\right)
\sum_{i} (-1)^{F}\int_{0}^{\infty} d {\rm{x}}\,{\rm{x}}^{2}\,{\rm{ln}}
\lt(1-(-1)^F {\rm {exp}}\{-\lt[{\rm{x}}^{2}+\frac{M_{i}^{2}(\phi)}{T^{2}}\rt]^{1/2} \} \rt)\,,
\eeq
where the sum is over all helicity states, $(-1)^{F}$ is $\pm1$ 
for bosonic and fermionic states, respectively, and $M_{i}(\phi)$ is the 
field-dependent mass of the {\it i}th state. For $\phi\ll T$ the temperature-dependent 
mass term is $\sigma \,T^{2}\lt|\phi\rt|^{2}$, where $\sigma \simeq 0.1$. 
For $T > T_c = M_s/\sqrt{\sigma}$ the potential
\beq \label{vt}
V(\phi)\,=\,\mu^4_0
\,+\,2(-M_{s}^{2}+\sigma T^{2})|\phi|^2
\,+\,2\,n^2\lambda^2\frac{\lt|\phi\rt|^{2(2n-1)}}{M_*^{2(2n-3)}}
\,+\,A_{\lambda} \lambda\frac{\lt(\phi\ov {\phi}\rt)^{n}}{M_*^{2n-3}}\,+\,{\rm c. c}\,,
\eeq
develops a minimum at $\lt|\phi\rt|=0$, the gauge group is unbroken and hence there are no monopoles. 
The false vacuum energy density $\mu_0^4$ ($\sim M_s^2 M^2$) drives thermal inflation and the universe 
experiences roughly ln$(\mu_0/T_c)$ $e$-foldings ($\sim 12$ for $M \sim 10^{14}$ GeV).

As the temperature $T$ falls below the critical value $T\sim T_c$,  
the mass-squared term for $\phi$ turns negative, and $\phi$ rolls from the origin to $M$, 
thereby ending thermal inflation. Monopoles are expected to arise as a 
consequence of symmetry breaking through 
the Kibble Mechanism \cite{Kibble:1976sj}. The field $\phi$ 
performs damped oscillations about the minimum at $M$ and subsequently 
decays. During these oscillations, the universe is matter dominated with energy
density $\rho \sim M_s^2 \la \phi \ra^2 (t^2_c/t^2)$, where $t_c$
represents the cosmic time at the phase transition. A large amount of entropy is 
released by the decay of $\phi$ field which helps dilute the 
initial monopole density \cite{Lazarides:1986rt}. 
To estimate the initial monopole number density $n_M$ one usually makes the plausible assumption
that a correlation size volume $\sim \xi^3 (\sim T_{c}^{-3})$ contains on the order of one
monopole  \cite{Kibble:1976sj, Zeldovich:1978wj}. Then  \cite{Kibble:1976sj, Zeldovich:1978wj}
\beq \label{nin}
r_{in} \equiv \lt[\frac{n_M}{T^3}\rt]_{initial} \sim \frac{P}{(4\pi/3)\,\xi^3\,T_c^3}\, 
\sim 10^{-2}\,,
\eeq 
where $P$ is a geometric factor of order $1/10$. 
It has been suggested \cite{Zurek:1996sj} that $\xi$ may be somewhat larger 
than $T_c ^{-1}$, in which case fewer monopoles would be produced. A 
more drastic reduction in the initial monopole number density is 
achieved by assuming that a single monopole is produced per horizon volume 
\cite{Einhorn:1980ym}. Since the horizon 
size during monopole production is larger than $T_c^{-1}$ by a factor $M_P/M$, 
a large suppression (by a factor of order $(M/M_P)^3)$ of the monopole 
number density becomes possible. We will not use this last suppression mechanism for superheavy 
monopoles, but will keep in mind the fact that the initial monopole number 
density estimates can have large uncertainties in them.

One may ask whether $r_{in}$ can be reduced through monopole-antimonopole 
annihilation \cite{Zeldovich:1978wj, Dicus:1982ri}. In the context of flaton models 
this has been investigated in \cite{Lazarides:1986rt} which reached the conclusion that 
there is no significant annihilation during the epoch of $\phi$ domination.

We now consider monopole dilution through 
entropy production from the decay of the flaton field. If the 
entropy increases by a factor $\Delta$, the final number density of monopoles 
is given by 
\beq \label{rfnal}
r_{final} \equiv \lt[\frac{n_M}{T^3}\rt]_{final}=r_{in}\Delta^{-1}\frac{g_*(T_{final})}{g_*(T_c)}\,,
\eeq
where $g_*(T)$ represents the degrees of freedom at $T$. 
The flaton decay proceeds predominantly via the coupling 
$\beta H_u H_d (\phi\ov {\phi})^{m}/{M_*^{2m-1}}$. The decay width $(\phi 
\rightarrow H_u^2,~H_d^2)$ is given by 
\beq
\Gamma_{\phi}\, \simeq \,\frac{1}{8 \pi}\beta^4 \lt(\frac{M}{M_*}\rt)^{4(2m-1)}\,\frac{M^2}{m_{\phi}}\,,
\eeq 
where $m_{\phi}=2\sqrt{2(n-1)}\,M_s$ is the flaton mass. The final temperature after thermal inflation can be 
expressed as 
\beq \label{tf}
T_f\,\simeq\,0.3\,\sqrt{\Gamma_{\phi} M_P}\simeq 0.036\,
\beta^2 \lt(\frac{M}{M_*}\rt)^{2(2m-1)} \frac{M}{\lt(n-1\rt)^
{1/4}}\,\sqrt{\frac{M_P}{M_s}}\,,
\eeq 
where $M_P=2.4\times 10^{18}$ GeV is the reduced Planck mass. Note that the last term in Eq. (\ref{spot}) 
generates the effective MSSM $\mu$ term $\beta (M/M_*)^{2m-1}M$. Eq. (\ref{tf}) can be rewritten as
\bea \label{tf1}
T_f &\simeq & 0.036\lt(\frac{\mu^2}{M}\rt) \frac{1}{\lt(n-1\rt)^{1/4}}\,\sqrt{\frac{M_P}{M_s}}\,,
\nonumber \\
& \simeq  & 0.01 \lt(\frac{\mu}{1\,{\rm TeV}}\rt)^2\,\lt(\frac{10^{14}\,{\rm GeV}}{M}\rt)\,
\lt(\frac{950 \,{\rm GeV}}{M_s}\rt)^{1/2}\,{\rm GeV}~~~({\rm for}\,n=4\,;m=3)\,.
\eea
The entropy released by flaton decay is estimated to be
\bea \label{ent}
\Delta &\simeq&\frac{3\mu_0^4}{g_*(T_c)T_c^3T_f}\,,\nonumber \\
&\simeq & 9 \times 10^{23}\, \lt(\frac{\sigma}{0.1}\rt)^{3/2}
\,\lt(\frac{M}{10^{14}\,{\rm GeV}}\rt)^3
\,\lt(\frac{1\,{\rm TeV}}{\mu}\rt)^2\,\lt(\frac{950\,{\rm GeV}}{M_s}\rt)^{1/2}\,,
\eea
where we have used Eq. (\ref{tf1}) and $g_*(T_c) \sim 200$. Using Eqs. (\ref{ent}) 
and (\ref{rfnal}) we get
\beq
r_{final} \,\simeq\,
6 \times 10^{-28} \,\lt(\frac{r_{in}}{10^{-2}}\rt)\,\lt(\frac{0.1}{\sigma}\rt)^{3/2}\,
\lt(\frac{10^{14}\,{\rm GeV}}{M}\rt)^3\,\lt(\frac{\mu}{1\,{\rm TeV}}\rt)^2 
\lt(\frac{M_s}{950\,{\rm GeV}}\rt)^{1/2}\,,
\eeq 
for $g_*(T_{f}) \sim 10$ \cite{Kolb:1990vq}. 

\begin{figure}[h]
\includegraphics[angle=270,width=8cm]{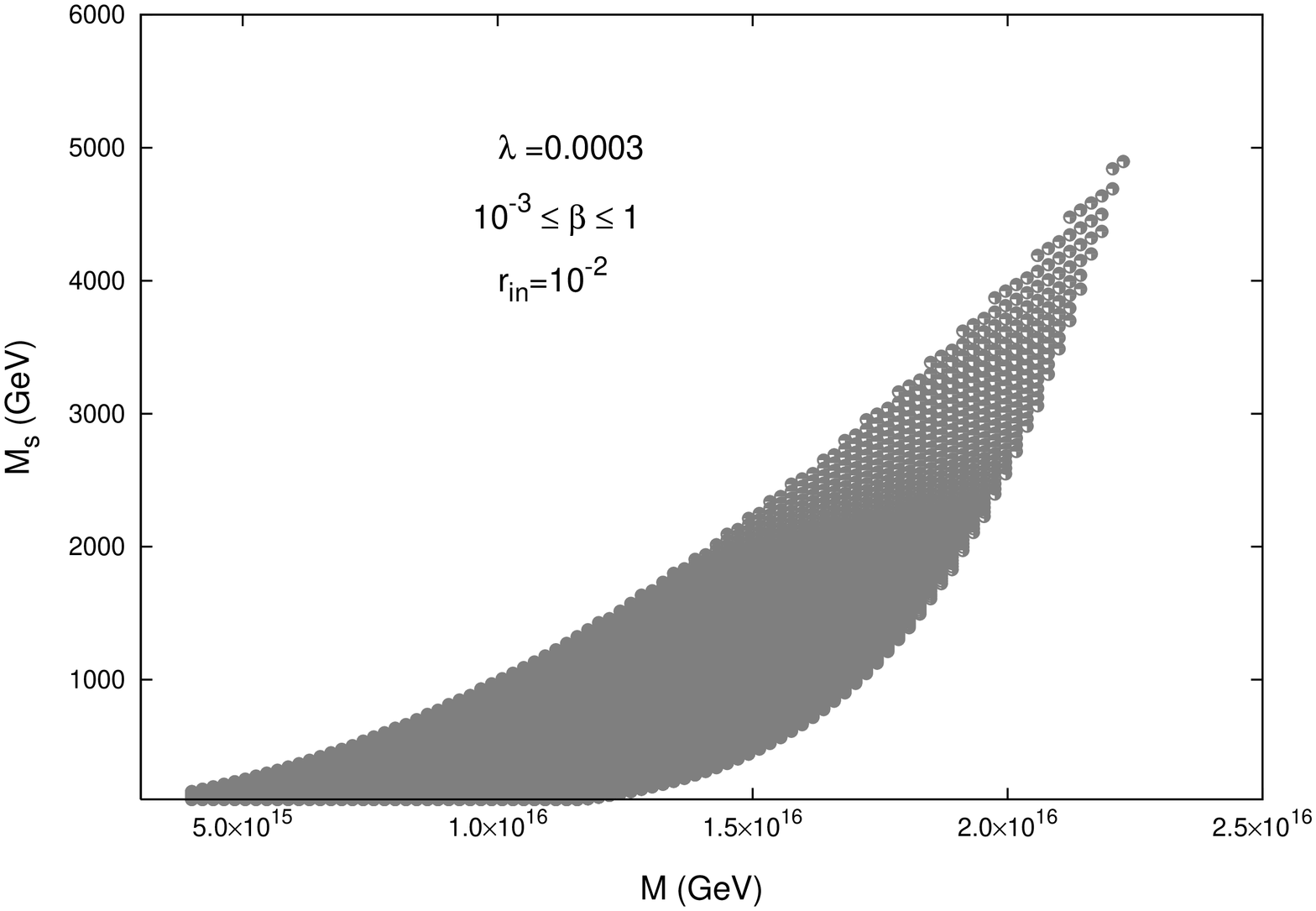}
\caption{$M_s$ vs $M$ (for superheavy monopoles), with
final monopole flux $F_M = 10^{-17}{\rm cm^{-2}}{\rm sec^{-1}}{\rm sr^{-1}}$.
Here 100 GeV $\le \mu \le$ 20 TeV.}
\label{fig1}
\end{figure}

From Eq. (\ref{tf1}) we note that if $M$ increases 
above $10^{15}$ GeV, the $\mu$ parameter and $M_s$ also have to increase
(see Fig. \ref{fig1}) in order that the final temperature 
$T_f$ stays above 10 MeV for successful nucleosynthesis. 
Thus, some fine tuning of the soft supersymmetry breaking 
higgs parameters will be required to implement electroweak breaking. Monopoles 
with masses greater than or of order $10^{17}$ GeV can gravitationally clump 
and will be considered later.

\begin{figure}[h]
\begin{tabular}{cc}
      \resizebox{70mm}{!}{\includegraphics{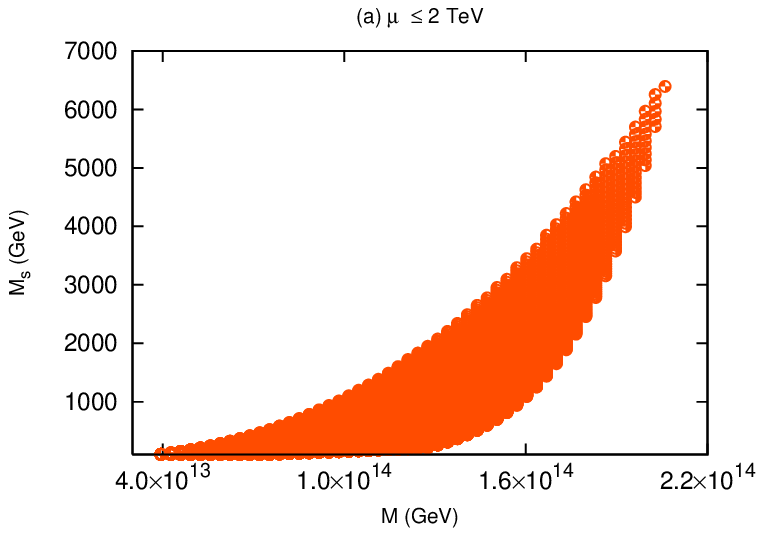}} &
      \resizebox{70mm}{!}{\includegraphics{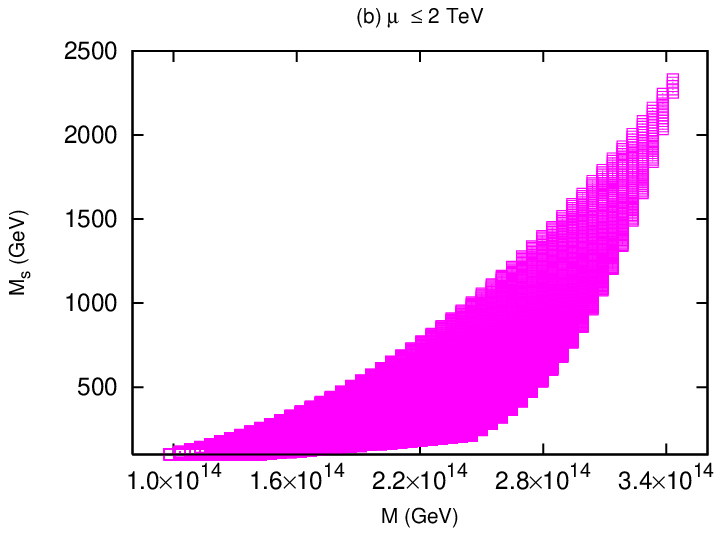}} \\
    \end{tabular}
    \caption{$M_s$ versus $M$ with $\lambda = 10^{-2}$ and $10^{-3} \le \beta \le 1$. Here 
$r_{in}=10^{-2}$. The monopole flux $F_M=10^{-17}{\rm cm^{-2}}{\rm sec^{-1}}{\rm sr^{-1}}$ 
for plot (a) and $10^{-18}{\rm cm^{-2}}{\rm sec^{-1}}{\rm sr^{-1}}$ for plot (b).}
\label{fig2}
\end{figure}

We should make sure that the final monopole number density is consistent with the MACRO 
bound \cite{Ambrosio:2002qq} which is more stringent than the well known 
Parker bound \cite{Lazarides:1980tf, Turner:1982ag}. 
For superheavy magnetic monopoles ($10^{11}$ GeV 
$\lesssim m_M \lesssim 10^{17}$ GeV) moving with speed 
$v_M \sim 3 \times 10^{-3}c \,(10^{16}\,{\rm GeV}/m_M)^{1/2}$ 
\cite{Lazarides:1980tf, Kolb:1990vq}, the MACRO limit on the flux $F_M$ is
\beq \label{fm}
F_M \lesssim \,10^{-16}{\rm cm^{-2}}{\rm sec^{-1}}{\rm sr^{-1}}\,,
\eeq
which corresponds to $r_{final} \lesssim 2 \times 10^{-26}
(3 \times 10^{-3} c/v_M)(m_M/10^{16}\,{\rm GeV})^{1/2}$. For $n=4, m=3$ in Eq. (\ref {spot}) and with 
the remaining parameters given in Table I (corresponding to $M=10^{14}$ GeV), an entropy release 
factor $\Delta$ of order $10^{22}$ saturates the bound in Eq. (\ref{fm}), 
with $r_{in} \sim 10^{-2}$ (see Eq. (\ref{nin})). The variation of $M_s$ and $M$ with the initial 
monopole number density given by Eq. (\ref{nin}), such that the monopole flux bound 
($F_M \simeq 10^{-17}{\rm cm^{-2}}{\rm sec^{-1}}{\rm sr^{-1}}$) 
is saturated, is shown in Fig. \ref{fig2}(a). For
$F_M= 10^{-18} {\rm cm^{-2}}{\rm sec^{-1}}{\rm sr^{-1}}$ $M_s$ vs. $M$ is plotted in 
Fig. \ref{fig2}(b).
\begin{table}[t]
\label{tb1}
\begin{tabular}{|c|c|c|c|c|c|c|c|c|}
\hline
$M$ & $M_s$ &$M_*$ & $T_f$ & $\mu$ &
$\lambda$ & $\beta$ & $r_{in}$ & $F_M$\\
(GeV)& (GeV)& (GeV)& (GeV)& (GeV)& & & &$({\rm cm^{-2}}{\rm sec^{-1}}{\rm sr^{-1}})$\\
\hline
\hline
$10^{14}$ & 950 & $10^{16}$ & $0.01$ & $10^3$ & $0.01$ & 0.1 & $10^{-2}$ &$10^{-17}$\\
\hline
$10^{15}$ & 3000 & $6.4 \times 10^{16}$ & 0.3 &$1.8 \times 10^4$ & 0.0003 & 0.02 & $10^{-2}$ & $10^{-18}$\\
\hline
$10^{16}$ & 3000& $10^{18}$ & 0.03 & $2 \times 10^{4}$ & 0.0003 & 0.02 & $10^{-2}$& $10^{-16}$\\
\hline
$2 \times 10^{16}$ & $2500$ & $2.4 \times 10^{18}$& 0.01& $1.6 \times 10^{4}$ & 
$0.0003$& 0.02& $10^{-2}$ & $10^{-17}$\\
\hline
\end{tabular}
\caption{\small A set of parameter values for which the expected monopole flux is at or below the 
MACRO bound.}
\end{table} 

The release of such a large amount of entropy (Eq. (\ref{ent})) 
certainly washes away any pre-existing baryon asymmetry. 
Also, for $M \gtrsim 10^{13}-10^{14}$ GeV 
the final temperature is quite low\footnote{$T_f$ as low as this can lead to conflict with 
LSP cosmology \cite{Moroi:1994rs}. One way out may be to introduce an axino LSP 
\cite{Rajagopal:1990yx}.} ($T_f \simeq 0.01$ GeV), so that 
the sphalerons are ineffective, and the standard leptogenesis scenario \cite{Fukugita:1986hr} 
does not apply. A different mechanism for 
generating the desired baryon asymmetry must then be found. 
This problem has arisen before and discussed 
by several authors \cite{Panagiotakopoulos:1987he, Dimopoulos:1987rk, 
Lazarides:1987yq}. 
In \cite{Lazarides:1987yq}, for instance, new particles beyond the MSSM spectrum are considered 
whose out of equilibrium decay can give rise to the 
baryon asymmetry. Here we rely on a modification of the 
AD \cite{Affleck:1984fy} scenario proposed in \cite{Stewart:1996ai} 
(and subsequently in \cite{Jeong:2004hy}) in which a dilution factor 
$\Delta^{-1} \sim 10^{-17}-10^{-18}$ and symmetry breaking 
scale $M \sim 10^{10-11}$ GeV are considered\footnote{For nonequilibrium effects in 
AD mechanism, see \cite{Charng:2005um}.}. 
In our case, as we saw in Eq. (\ref{ent}), the monopole problem 
requires an even greater amount of entropy production, especially if 
$r_{in}\sim 10^{-2}$. 

To implement the scenario discussed in \cite{Stewart:1996ai} in our case, we couple $\phi$ to the 
right-handed neutrino superfields $N_i$. In the MSSM notation, consider the couplings\footnote{
In practice, the superpotential is invariant under $G$, and $N_i$ belong in some appropriate 
$G$ representation. Later we will provide an example for $G$ and discuss how the terms in Eq. 
(\ref{spot1}) arise.}
\beq \label{spot1}
W\,=\,W_{inf}
\,+\, Y_l L H_d e
\,+\,Y_D L H_u N
\,+\, \lambda_{\phi} \phi N^2+{}\cdots\,,
\eeq
where we have suppressed all generation and group indices, and the ellipsis 
represent terms in the superpotential which will not participate in the analysis. 
We assume that both during and after thermal inflation, 
the squark fields do not acquire non-zero vevs. 

Consider the zero temperature $F$-term potential
\bea \label{Fpot}
V_F&=& \lt|Y_{D} H_u N + Y_l H_d e\rt|^2
\,+\,\lt|Y_{D} L H_u+ 2 \lambda_{\phi}\phi N\rt|^2
\,+\,\lt|Y_l L e +\beta H_u \frac{\lt(\phi \ov{\phi}\rt)^3}{M_*^5}\rt|^2 \nonumber \\
&+&\lt|Y_{D} L N + \beta H_d \frac{\lt(\phi \ov{\phi}\rt)^3}{M_*^5}\rt|^2
\,+\,\lt| \frac{4 \lambda \lt(\phi \ov{\phi}\rt)^3}{M_*^5}\phi
+3 \beta H_u H_d \frac{\lt(\phi \ov{\phi}\rt)^2}{M_*^5}\phi\rt|^2\nonumber \\
&+&\lt| \frac{4 \lambda \lt(\phi \ov{\phi}\rt)^3}{M_*^5}{\ov \phi}
+\lambda_{\phi}N^2
+3 \beta H_u H_d \frac{\lt(\phi \ov{\phi}\rt)^2}{M_*^5}{\ov \phi}\rt|^2,
\eea
where we have used the same notation for superfields and their scalar components and set $n=4, m=3$ 
in Eq. (\ref{spot}). 
The $D$-terms ($V_D$) and the supersymmetry breaking parts of the potential ($V_{\not \!\! {susy}}$) are
\bea
V_D&+& V_{\not \!\! {susy}}=g^2 \lt(\lt|H_u\rt|^2-\lt|H_d\rt|^2-\lt|L\rt|^2\rt)
\,+\,m_{L}^2 \lt|L\rt|^2
\,-\,m_{H_u}^2 \lt|H_u\rt|^2
\,-\,2M_{s}^2 \lt|\phi\rt|^2 \nonumber \\
&+&\lt(A_{D} Y_{D}L H_u N+A_{\lambda_{\phi}} \lambda_{\phi} \phi N^2
+A_{\beta} \beta H_u H_d \frac{\lt(\phi \ov{\phi}\rt)^3}{M_*^5}
+A_{\lambda} \lambda \frac{\lt(\phi \ov{\phi}\rt)^4}{M_*^5} +{}\cdots \rt)+ {\rm c. c}\,,
\eea
where the soft masses and $A$-terms are all taken to be of order $M_s$, 
and $D$-flatness along $\phi, {\ov \phi}$ 
direction has been taken into account. For simplicity we will confine ourselves to 
a single generation picture. 

We assume that initially all fields are held at zero during thermal inflation. To implement 
the scenario, the AD-field, parameterized along the $D$-flat direction by $LH_u=\psi^2/2$, 
acquires a vev $\sim m_{\psi}^2/\lt|Y_{D}\rt|^2$ provided we set 
$m_{\psi}^2=(m_{H_u}^2-m_L^2)/2 > 0$ ($\phi$ is still at zero). 
Assuming this happens before the flaton field starts 
to roll down, the term $A_D Y_D L H_u N+{\rm c.c.}$ induces a 
vev for $N$ which, in turn, triggers 
the slow roll of $\phi$ towards its minimum. As $\phi$ increases, $N$ follows its instantaneous minimum
\beq \label{nvev}
N\,\simeq \,-\frac{Y_{D} L H_u}{2 \lambda_{\phi}\phi}\,.
\eeq 
Using Eq. (\ref{nvev}) the effective potential reduces to\footnote{For simplicity $e$ and $H_d$ 
are assumed to be at their zero location.}
\bea \label{psipot}
V &=& V_F+V_D + V_{\not \!\! susy}\,,\nonumber \\
&=& 2 \lt| \frac{4 \lambda}{M_*^5}\rt|^2 \lt|{\phi}\rt|^{14}
\,-\, 2M_{s}^2\lt|\phi\rt|^2
\,+\,\lt(\frac{1}{2}\lt|\frac{\beta}{M_*^5}\rt|^2 \lt|\phi\rt|^{12}
\,-\, m_{\psi}^2\rt) \lt|\psi\rt|^2
\,+\, \lt|\frac{Y_{D}^2 \psi^3}{4 \lambda_{\phi} \phi}\rt|^2 \nonumber \\
&+& \lt[\lt\{\lt(\frac{A_{\lambda_{\phi}}}{2}-A_{D}+ 2\lambda^* \frac{\lt(\phi {\ov \phi}\rt)^3 {\ov \phi}^*}
{M_*^5 \phi}\rt)\frac{Y_{D}^2 \psi^4}
{8 \lambda_{\phi} \phi}
\,+\, A_{\lambda} \lambda \frac{\lt (\phi {\ov \phi} \rt)^4}{M_*^5}\rt\}
\,+\, {\rm c. c}\rt]\,.
\eea
It follows, after some algebra, that for 
$\phi\ll M$, $\psi$ is given by
\beq \label{psi}
\lt|\psi\rt|^2 \simeq 4 \sqrt{\frac{1}{3}} \lt(\frac{\lambda_{\phi}}{Y_D^2}\rt) M_s\lt|\phi\rt|\,,
\eeq
and its phase is related with that of $\phi$ from the term
\beq \label{phase}
\lt(\frac{A_{\lambda_{\phi}}}{2}-A_{D}+ 2\lambda^* \frac{\lt(\phi {\ov \phi}\rt)^3 
{\ov \phi}^* }{M_*^5 \phi}\rt) + {\rm c. c}\,.
\eeq
Note that at this stage the third term in Eq. (\ref{phase}) is negligible.

When $\phi$ becomes comparable to $M$ (with the temperature still around $T_c$), 
it yields a positive mass squared term for $\psi$, provided that
\beq 
\frac{1}{2} \beta^2 \frac{|\phi|^{12}}{M_*^{10}} - m_{\psi}^2 >0 \,.
\eeq 
This helps the $LH_u$ direction  to reach its true minimum at zero\footnote{From the tabulated 
values of $\lambda$, $\beta$ (corresponding to $M=10^{14}$ GeV) and Eq. (\ref{nbs}) we notice that this 
happens before $\phi$ actually reaches its minimum (as $\lambda \ll \beta$).}. 
Simultaneously, the phase of $\psi$ gets an important contribution from the third term inside the bracket 
of Eq. (\ref {phase}). This change then initiates an angular momentum in $LH_u$ which 
turns out to be the amount of lepton asymmetry $(n_L=i(\psi^* \dot{\psi}- \psi \dot{\psi}^*))$ 
generated from the AD mechanism. The difference between $(A_{\lambda_{\phi}}/2M_s-A_D/M_s)$ and 
$2 \lambda^*$ would be the source of $CP$-violation.

To protect $n_L$ from being erased by potential lepton-number violating processes, 
in Ref. \cite{Stewart:1996ai} it is assumed that the AD field 
should completely decay while the universe is still dominated by the energy of the 
flaton field. We assume that the decay width of the AD field is such that the electroweak sphalerons 
\cite{Kuzmin:1985mm} can convert a fraction of $n_L$ into $n_B$, the baryon number density. 

The decay of $\phi$ yields a final temperature of around 
$0.01$ GeV. The release of entropy (estimated earlier) responsible 
for the dilution of monopoles also dilutes the baryon asymmetry. The final 
baryon asymmetry is approximately given by 
\bea \label{nbs}
\frac{n_B}{s} &\simeq& \frac{1}{3}\,\frac{n_L}
{s_{in}}\,\Delta^{-1} \simeq \frac{n_L}{s_{final}}\,,
\nonumber \\
&\sim & \frac{n_L}{\rho_{\phi}}\,T_f\,\sim\,\frac{4}{3}\,\sqrt{\frac{1}{3}}\,\frac{\lambda_{\phi}}{Y_D^2}\,
\frac{T_f}{M}\,\frac{m_{\psi}}{M_s}\frac{n_L}{n_{\psi}}\,,
\eea  
where we have used Eq. (\ref{psi}). The seesaw 
relation for the light neutrino mass is \footnote
{From Eq. (\ref{spot1}), it is seen that the effective superpotential responsible for
generating the neutrino mass is $W_{\nu} = - (Y_D L H_u)^2/(2 \lambda_{\phi} \phi)$ after integrating out $N$.}
\beq \label{27}
m_{\nu}\simeq \frac{Y_{D}^2 v^2}{\lambda_{\phi} M},
\eeq
where we employ a basis in which the light neutrino mass matrix
is diagonal and, for simplicity, only a single $LH_u$ flat direction is assumed. 
Substituting Eq. (\ref{27}) in Eq. (\ref{nbs}), 
\beq\label{26}
\frac{n_B}{s} \sim \frac{4}{3}\,\sqrt{\frac{1}{3}}\,\frac{v^2}{m_{\nu} M^2}\,\frac{m_{\psi} T_f}{M_{s}}\,,
\eeq
where $n_L/n_{\psi} \sim {\cal O}(1)$ is assumed.
 
\begin{figure}[h]
\includegraphics[angle=270,width=8cm]{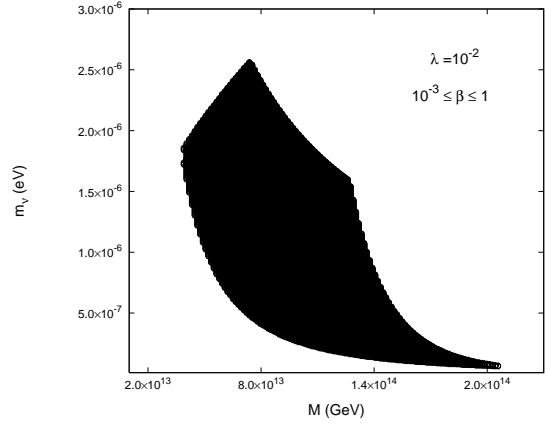}
\caption{$m_{\nu}$ versus $M$ for $F_M=10^{-17}{\rm cm^{-2}}{\rm sec^{-1}}{\rm sr^{-1}}$
and $r_{in}=10^{-2}$
(corresponding to Fig. \ref{fig2} (a)), with $n_B/s \sim (0.837- 0.895)\times 10^{-10}$ \cite{Spergel:2006hy}.}
\label{fig3}
\end{figure}

To obtain the required $n_B/s$, the neutrino mass $m_{\nu}$ in Eq. (\ref{26}) turns 
out to be several orders of magnitude below
the scale for atmospheric and solar neutrino oscillations. From Eq. (\ref{26}),
\beq \label{ns}
\frac{n_B}{s}\,\sim \, 10^{-10} \lt(\frac{v}{174\,
{\rm GeV}}\rt)^2\lt(\frac{2 \times 10^{-7}\, {\rm eV}}{m_{\nu}}\rt)
\lt(\frac{10^{14}\, {\rm GeV}}{M}\rt)^2\,\lt(\frac{T_f}{0.01\, {\rm GeV}}\rt)\,,
\eeq
where we have set $m_{\psi}\sim M_s$, and
from the observed baryon to photon ratio $n_B/n_{\gamma} \simeq 
(6.0965 \pm  0.2055) \times 10^{-10}$ \cite{Spergel:2006hy}, $n_B/s \simeq 
(n_B/n_{\gamma})/7.04$ \cite{Kolb:1990vq}. In Fig. \ref{fig3} we display the allowed region for
$m_{\nu}$, with the range of $M$ and $T_f$ restricted by the monopole flux corresponding
to Fig. \ref{fig2} (a).

In order to make the discussion more explicit and to realize a scenario 
with superheavy symmetry breaking scale corresponding to $n=4$, $m=3$, we 
consider a realistic model with gauge symmetry $G_{422}\equiv SU(4)_c \times 
SU(2)_L \times SU(2)_R$ \cite{Salam:1973uk}, whose breaking yields monopoles which carry 
two quanta of Dirac magnetic charge \cite{Lazarides:1980cc}. By the 
same token, in principle, there also can exist color 
singlet states with charge $\pm e/2$ \cite{King:1997ia, Kephart:2006zd}\footnote{If $G_{422}$ 
is replaced by $G_{333} \equiv SU(3)_c \times SU(3)_L \times SU(3)_R$ we 
find monopoles carrying three quanta of Dirac charge \cite{Shafi:1984wk}.}. 
We have checked that $M$ as low as $10^{14}$ GeV is compatible with proton lifetime limits. Indeed, 
for $M \sim 10^{14}$ GeV, we estimate the proton lifetime to be of order 
$10^{34}$ yrs, taking into account the operators discussed in \cite{Babu:1997js}. 
The quarks and leptons are unified in the representations 
$F_i = (4,2,1)_i$; $\ov{F}_i =(\ov{4},1,\ov{2})_i$ of $G_{422}$, 
where the subscript $i\ (=1,2,3)$ denotes the family index. 
The Higgs sector consists of 
\beq
h=(1,2,\bar{2})\,;~~~{\ov{\phi}}  = (\ov{4},1,\ov{2})\,;~~~{\phi} =  (4,1,2)\,.
\eeq
The superpotential $W$ is given by 
\bea \label{spot2}
W &=& \lambda\frac{\lt(\phi\ov {\phi}\rt)^{4}}{M_*^{5}}
\,+\,\beta\,h\,h\,\frac{\lt(\phi\ov {\phi}\rt)^{3}}{M_*^{5}}
\,+\,Y F h {\ov F}
\,+\,\gamma \,\phi \, \phi \frac{{\ov F}\,{\ov F}}{M_*}
\,+\,\kappa \,{\ov \phi}\,{\ov \phi} \frac{{\ov F}\, {\ov F}}{M_*} \nonumber \\
&+& a_1 {\cal D}\phi \phi
\,+\,a_2 {\cal D}\,{\ov \phi}\,{\ov \phi}
\,+\,\frac{a_3}{M_*}{\cal D} {\cal D} \lt(\frac{\phi {\ov \phi}}{M_*}\rt)^2\
\,+\, a_4 F F {\cal D} \frac{\phi {\ov \phi}}{M_*^2}\,+{}\cdots\,,
\eea
where a discrete symmetry $Z_4 \times Z_8$ (see Table II) has been introduced to realize 
the desired $D$ and $F$-flat direction\footnote{To avoid topologically stable 
domain walls we will assume that the discrete symmetry is explicitly broken 
by higher dimensional operators.}. 
The color sextet superfield ${\cal D}=(6, 1, 1)=(D^c, \ov{D}^c)$, where $D^c=(3, 1, 1/3)$ and
$\ov{D}^c=(\ov{3}, 1, -1/3)$, is introduced to provide heavy mass to the components 
$d_H^c, \ov{d}_H^c$ of $\phi, \ov{\phi}$ \cite{Antoniadis:1988cm, King:1997ia}. 

\begin{table}[t]
\begin{center}
\begin{tabular}{|c|c|c|c|c|c|c|}
\hline
Fields & $\phi$ & $\overline{\phi}$ & $h$ & $F$ & $\ov{F}$ & ${\cal D}$ \\
\hline
\hline
$Z_4$ & 1 & -1 & $i$ & $i$ & -1 & 1 \\
\hline
$Z_8$ & $\omega$ & $\omega$ & $\omega$ & $\omega^4$ & $\omega^3$ & $\omega^6$ \\
\hline
\end{tabular}
\end{center}
\caption{\small Discrete charges of various superfields.}
\end{table}
The interaction of $\phi$ with $N$ 
in this particular example is determined by the term $(\gamma \phi^2
+\kappa \bar \phi^2 )N^2/M_*$ in the superpotential 
instead of $\lambda_{\phi} \phi N^2$ in Eq. (\ref{spot1}). 
Thus as long as $\phi$ is held at zero, $N$ does not acquire 
any vev through the term $A Y F h {\ov F}$ 
and from $V_F$ even when the AD field has a vev\footnote{The AD field, $\psi$, 
in this example is a flat direction chosen along the neutral 
components of $F$ and $h$ ($H_u^o$).}. 
The $\phi$ field starts to roll down 
when $T \sim M_s$ as discussed in the context of 
Eq. (\ref{vt}). As $\phi$ increases, $N$ would receive a vev $\sim (Y \la \psi
\ra^2/4 (\gamma+\kappa)\lt|\phi\rt|)(M_*/\lt|\phi\rt|)$ similar to 
Eq. (\ref{nvev}), 
with $\lambda_{\phi} \sim (\gamma+\kappa) (\lt|\phi\rt|/M_*)=\gamma^{'}(\lt|\phi\rt|/M_*)$. 
The relation in Eq. (\ref{psi}) now becomes
\beq \label{psi1}
\lt|\psi\rt|^2 \simeq 4 \sqrt{\frac{1}{3}} \lt(\frac{\gamma^{'}}{Y_D^2}\rt)\frac{M_sM}{M_*}\lt|\phi\rt| \,.
\eeq
The rest of the discussion for an estimate of the observed baryon asymmetry is very similar 
to what we already have before Eq. (\ref{ns}) .

The baryogenesis scenario 
we have discussed, following \cite{Stewart:1996ai}, can also be employed for 
monopoles with masses $\sim 5 \times 10^{17}$ GeV, corresponding to GUT symmetry breaking scale 
$M \sim 2 \times 10^{16}$ GeV. 
The gravitational force on such monopoles exceeds the magnetic force 
and it seems plausible that they would clump in the galaxy, and perhaps even 
contribute to the dark matter in the universe. (This may not be plausible for 
monopoles that catalyze nucleon decay). However, this latter 
possibility is disfavored by the MACRO bound \cite{Witten:2002wb}. To see this, assume that 
the monopole energy density $\rho_M \lesssim \rho_{B}$, the baryon energy density. 
Taking a local density enhancement factor of order $10^5$, the local flux is estimated 
to be (with $v_M=10^{-3}c$)
\begin{equation}
F_M \lesssim 6 \times 10^{-13}\,\lt(\frac{10^{17} {\rm{GeV}}}{m_M}\rt)\, 
{\rm cm^{-2}}{\rm sec^{-1}}{\rm sr^{-1}};\,\,\,m_M \sim 10^{17}-10^{18} 
{\rm GeV}\,,
\end{equation}
which is in strong disagreement with the MACRO bound. Thus, 
$\rho_M \lesssim 10^{-4}\rho_B$, to be consistent with the MACRO bound. 
To achieve a monopole flux $\lesssim 10^{-17}{\rm cm^{-2}}{\rm sec^{-1}}
{\rm sr^{-1}}$, we estimate that $r_{final} \lesssim 10^{-31}$. 
The parameters chosen to achieve this are displayed in Table I. 

\begin{figure}[h]
\includegraphics[angle=270,width=8cm]{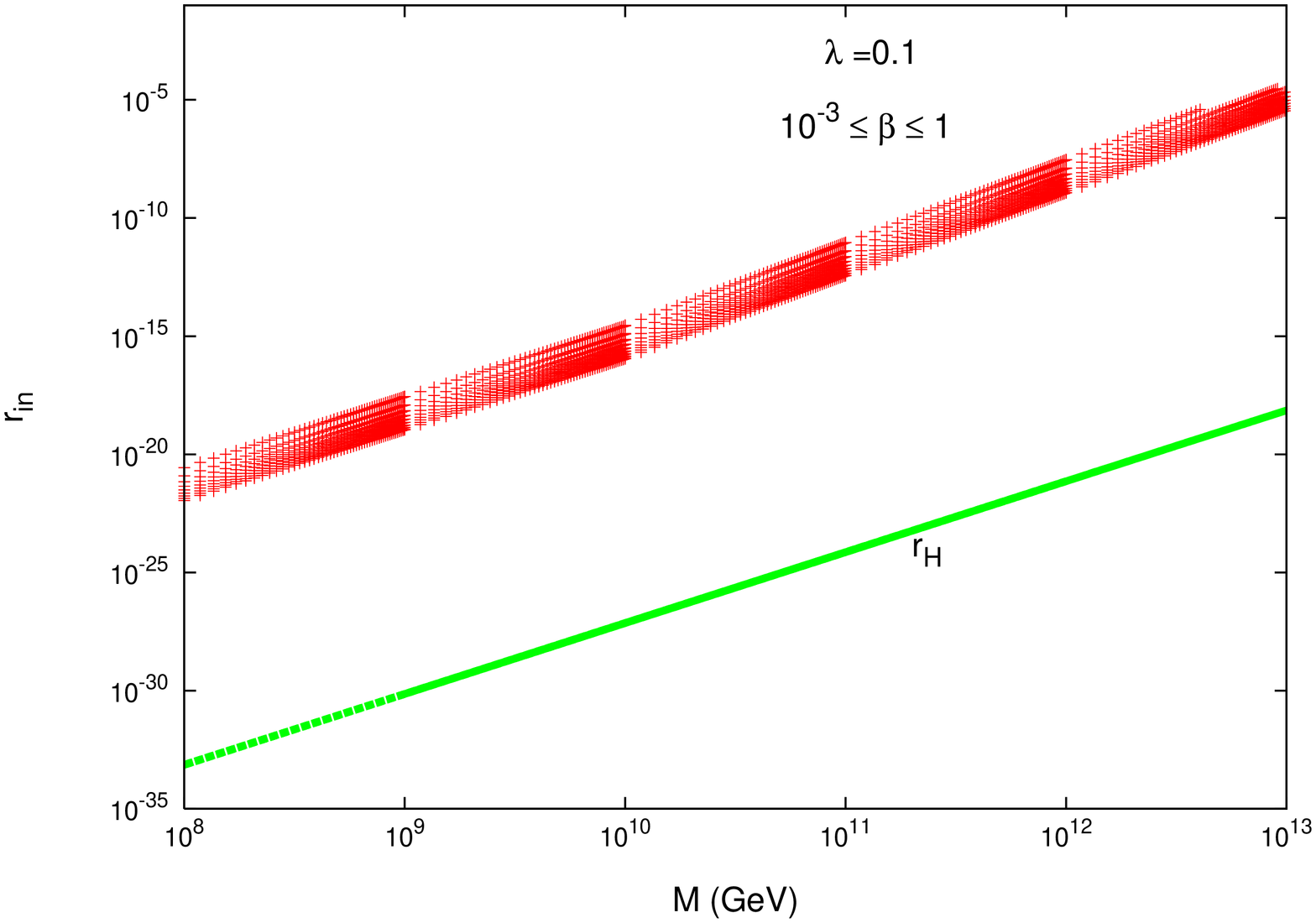}
\caption{$r_{in}$ vs. $M$ (for relativistic monopoles), 
with $F_M = 10^{-17}{\rm cm^{-2}}{\rm sec^{-1}}{\rm sr^{-1}}$, 
100 GeV $\le M_s \le$ 1 TeV and 100 GeV $\le \mu \le$ 2 TeV. 
$r_H$ represents the relative monopole number density assuming production of one monopole per 
horizon volume.}
\label{fig4}
\end{figure}

Let us now consider `lighter' monopoles with masses of order 
$10^9-10^{10}$ GeV, corresponding to symmetry breaking scales $\sim 10^8-10^9$ GeV. (In models with 
$D$ and $F$-flat directions the symmetry breaking scale is expected to be $\gtrsim 10^{8}$ GeV). 
In this case the amount of entropy released following thermal inflation is 
considerably smaller. Namely, from Eq. (\ref{ent}), $\Delta \sim 10^6$, so that a 
monopole flux of order $10^{-17}\,{\rm cm^{-2}}{\rm sec^{-1}}{\rm sr^{-1}}$ (this is 
consistent with present bounds from MACRO \cite{Ambrosio:2002qq}, SLIM \cite{Balestra:2006fr} 
and AMANDA \cite{Niessen:2001ci} experiments) requires that the 
initial relative monopole number density $r_{in}$ must be sufficiently small, 
as shown in Fig. \ref{fig4} (see discussion following Eq. (\ref{nin})). 
Following \cite{Dar:2003cr}, 
the observed baryon asymmetry can now 
be explained via TeV scale leptogenesis, since the final temperature $T_f \sim$ TeV. 
An initially 
large lepton asymmetry can survive the moderate amount of 
dilution so as to produce a final baryon asymmetry 
consistent with observations. 
Finally, we see from Fig. \ref{fig4}, 
that a flux bound of intermediate mass monopoles ($\sim10^{11}- 10^{13}$ GeV) 
requires that $r_{in} \sim 10^{-17} - 10^{-7}$. 
The baryon asymmetry here can be achieved either through TeV scale leptogenesis 
\cite{Dar:2005hm} or the original AD scenario \cite{Affleck:1984fy} where $T_f$ is of order 100 GeV. For monopoles of mass 
$\sim 10^{13}- 10^{14}$ GeV, the baryon asymmetry can be generated 
through the modified Affleck-Dine mechanism.

To summarize, magnetic monopoles appear in a variety of unified gauge models 
with a wide range of masses. In models of thermal inflation, monopoles with masses of order 
$10^9-10^{18}$ GeV appear when the cosmic 
temperature is in the TeV range. Their subsequent dilution through the release 
of huge amount of entropy poses a challenge for baryogenesis. We have discussed 
a class of realistic models in which the observed baryon asymmetry can co-exist 
with a primordial monopole flux which can be detected with large 
scale detectors such as ICE CUBE \cite{Spiering:2005xv}.\\
 
The authors thank K. S. Babu, Daniel Chung, Ilia Gogoladze,  K. Hamaguchi, C. N. Leung, 
Peter Niessen, Arvid Pohl, V. N.  {\c S}eno$\breve{\textrm{g}}$uz, Zurab Tavarkiladze and 
Henrike Wissing for many valuable discussions. 
This work was supported by the DOE under Contract No. DE-FG02-91ER40626.

\begin{thebibliography}{xx}

\bibitem{Salam:1973uk} 
J.~C.~Pati and A.~Salam, Phys.\ Rev.\ D {\bf 10} (1974) 275.

\bibitem{Georgi:1974sy}
H.~Georgi and S.~L.~Glashow,
Phys.\ Rev.\ Lett.\  {\bf 32}, 438 (1974); 
H.~Georgi, H.~R.~Quinn and S.~Weinberg,
Phys.\ Rev.\ Lett.\  {\bf 33}, 451 (1974).

\bibitem{Polyakov:1974ek}
A.~M.~Polyakov,
JETP Lett.\  {\bf 20} (1974) 194
[Pisma Zh.\ Eksp.\ Teor.\ Fiz.\  {\bf 20} (1974) 430]; G.~'t Hooft,
Nucl.\ Phys.\ B {\bf 79}, 276 (1974).

\bibitem{Daniel:1979yz}
M.~Daniel, G.~Lazarides and Q.~Shafi,
Nucl.\ Phys.\ B {\bf 170}, 156 (1980); 
C.~P.~Dokos and T.~N.~Tomaras,
Phys.\ Rev.\ D {\bf 21}, 2940 (1980).

\bibitem{Rubakov:1981rg}
V.~A.~Rubakov,
JETP Lett.\  {\bf 33}, 644 (1981)
[Pisma Zh.\ Eksp.\ Teor.\ Fiz.\  {\bf 33}, 658 (1981)]; 
Nucl.\ Phys.\ B {\bf 203}, 311 (1982); C.~G.~.~Callan,
Phys.\ Rev.\ D {\bf 25}, 2141 (1982).

\bibitem{Lazarides:1980cc}
G.~Lazarides, M.~Magg and Q.~Shafi,
Phys.\ Lett.\ B {\bf 97}, 87 (1980).

\bibitem{Shafi:1984wk}
Q.~Shafi, in {\it Proceedings of the NATO Advanced Study Institute on 
Monopole '83, Ann Arbor, Michigan, 1983}, edited by J.~L.~Stone, NATO ASI Series B, vol III (Plenum, 
New York, 1984), p. 47.

\bibitem{Wen:1985qj}
X.~G.~Wen and E.~Witten,
Nucl.\ Phys.\ B {\bf 261}, 651 (1985).

\bibitem{Witten:2002wb}
E.~Witten,
[arXiv:hep-th/0212247].

\bibitem{Leontaris:2000hh}
G.~K.~Leontaris and J.~Rizos,
Phys.\ Lett.\ B {\bf 510}, 295 (2001)
[arXiv:hep-ph/0012255]; G.~K.~Leontaris and J.~Rizos,
Phys.\ Lett.\ B {\bf 632}, 710 (2006)
[arXiv:hep-ph/0510230].

\bibitem{Candelas:1985en}
P.~Candelas, G.~T.~Horowitz, A.~Strominger and E.~Witten,
Nucl.\ Phys.\ B {\bf 258}, 46 (1985); E.~Witten,
Nucl.\ Phys.\ B {\bf 258}, 75 (1985).

\bibitem{Prikas:2003ce}
A.~Prikas and N.~D.~Tracas,
New J.\ Phys.\  {\bf 5}, 144 (2003)
[arXiv:hep-ph/0303258].

\bibitem{Aldazabal:2000cn}
G.~Aldazabal, S.~Franco, L.~E.~Ibanez, R.~Rabadan and A.~M.~Uranga,
JHEP {\bf 0102}, 047 (2001)
[arXiv:hep-ph/0011132]; L.~L.~Everett, G.~L.~Kane, S.~F.~King, S.~Rigolin and L.~T.~Wang,
Phys.\ Lett.\ B {\bf 531}, 263 (2002)
[arXiv:hep-ph/0202100].

\bibitem{Zeldovich:1978wj}
Y.~B.~Zeldovich and M.~Y.~Khlopov,
Phys.\ Lett.\ B {\bf 79} (1978) 239; J.~Preskill,
Phys.\ Rev.\ Lett.\  {\bf 43}, 1365 (1979).

\bibitem{Guth:1980zm}
A.~H.~Guth,
Phys.\ Rev.\ D {\bf 23}, 347 (1981). 

\bibitem{Linde:1981mu}
A.~D.~Linde,
Phys.\ Lett.\ B {\bf 108}, 389 (1982); A.~Albrecht and P.~J.~Steinhardt,
Phys.\ Rev.\ Lett.\  {\bf 48}, 1220 (1982).


\bibitem{Turner:1982kh}
M.~S.~Turner,
Phys.\ Lett.\ B {\bf 115}, 95 (1982); G.~Lazarides, Q.~Shafi and W.~P.~Trower,
Phys.\ Rev.\ Lett.\  {\bf 49}, 1756 (1982).

\bibitem{Shafi:1984tt}
G. Lazarides and Q.~Shafi,
Phys.\ Lett.\ B {\bf 148}, 35 (1984).

\bibitem{Shafi:2006cs}
Q.~Shafi and V.~N.~Senoguz,
Phys.\ Rev.\ D {\bf 73}, 127301 (2006)
[arXiv:astro-ph/0603830].


\bibitem{Parker:1970xv}
E.~N.~Parker,
Astrophys.\ J.\  {\bf 160}, 383 (1970); Astrophys.\ J.\  {\bf 163}, 255 (1971);
Astrophys.\ J.\  {\bf 166}, 295 (1971).

\bibitem{Ambrosio:2002qq}
M.~Ambrosio {\it et al.}  [MACRO Collaboration],
Eur.\ Phys.\ J.\ C {\bf 25}, 511 (2002)
[arXiv:hep-ex/0207020]; G.~Giacomelli  [MACRO Collaboration],
[arXiv:hep-ex/0210021]; G.~Giacomelli and L.~Patrizii,
[arXiv:hep-ex/0506014].

\bibitem{Lazarides:1993fi}
G.~Lazarides and Q.~Shafi,
Phys.\ Lett.\ B {\bf 308}, 17 (1993)
[arXiv:hep-ph/9304247].

\bibitem{Langacker:1980kd}
P.~Langacker and S.~Y.~Pi,
Phys.\ Rev.\ Lett.\  {\bf 45}, 1 (1980).

\bibitem{Lazarides:1980va}
G.~Lazarides and Q.~Shafi,
Phys.\ Lett.\ B {\bf 94}, 149 (1980).

\bibitem{Dvali:1995cj}
G.~R.~Dvali, A.~Melfo and G.~Senjanovic,
Phys.\ Rev.\ Lett.\  {\bf 75}, 4559 (1995)
[arXiv:hep-ph/9507230].

\bibitem{Dvali:1997sa}
G.~R.~Dvali, H.~Liu and T.~Vachaspati,
Phys.\ Rev.\ Lett.\  {\bf 80}, 2281 (1998)
[arXiv:hep-ph/9710301].

\bibitem{Stojkovic:2004hz}
D.~Stojkovic and K.~Freese,
Phys.\ Lett.\ B {\bf 606}, 251 (2005)
[arXiv:hep-ph/0403248].

\bibitem{Lazarides:1985bj}
G.~Lazarides, C.~Panagiotakopoulos and Q.~Shafi,
Phys.\ Rev.\ Lett.\  {\bf 56}, 432 (1986).

\bibitem{Lazarides:1986rt}
G.~Lazarides, C.~Panagiotakopoulos and Q.~Shafi,
Phys.\ Rev.\ Lett.\  {\bf 58}, 1707 (1987).

\bibitem{Lazarides:1985ja}
G.~Lazarides, C.~Panagiotakopoulos and Q.~Shafi,
Phys.\ Rev.\ Lett.\  {\bf 56}, 557 (1986).

\bibitem{Yamamoto:1985rd}
K.~Yamamoto,
Phys.\ Lett.\ B {\bf 168}, 341 (1986).

\bibitem{Binetruy:1986ss}
P.~Binetruy and M.~K.~Gaillard,
Phys.\ Rev.\ D {\bf 34} (1986) 3069.

\bibitem{Lazarides:1987yq}
G.~Lazarides, C.~Panagiotakopoulos and Q.~Shafi,
Nucl.\ Phys.\ B {\bf 307}, 937 (1988).

\bibitem{Lazarides:1992gg}
G.~Lazarides and Q.~Shafi,
Nucl.\ Phys.\ B {\bf 392}, 61 (1993).

\bibitem{Lyth:1995ka}
D.~H.~Lyth and E.~D.~Stewart,
Phys.\ Rev.\ D {\bf 53}, 1784 (1996)
[arXiv:hep-ph/9510204].

\bibitem{Barreiro:1996dx}
T.~Barreiro, E.~J.~Copeland, D.~H.~Lyth and T.~Prokopec,
Phys.\ Rev.\ D {\bf 54}, 1379 (1996)
[arXiv:hep-ph/9602263].

\bibitem{Stewart:1996ai}
E.~D.~Stewart, M.~Kawasaki and T.~Yanagida,
Phys.\ Rev.\ D {\bf 54}, 6032 (1996)
[arXiv:hep-ph/9603324].

\bibitem{Kephart:2006zd}
T.~W.~Kephart and Q.~Shafi,
Phys.\ Lett.\ B {\bf 520}, 313 (2001)
[arXiv:hep-ph/0105237]; 
T.~W.~Kephart, C.~A.~Lee and Q.~Shafi,
[arXiv:hep-ph/0602055].

\bibitem{Jeannerot:1998qm}
R.~Jeannerot,
Phys.\ Rev.\ D {\bf 59}, 083501 (1999)
[arXiv:hep-ph/9808260].

\bibitem{Dar:2003cr}
S.~Dar, S.~Huber, V.~N.~Senoguz and Q.~Shafi,
Phys.\ Rev.\ D {\bf 69}, 077701 (2004)
[arXiv:hep-ph/0311129].

\bibitem{Dar:2005hm}
S.~Dar, Q.~Shafi and A.~Sil,
Phys.\ Lett.\ B {\bf 632}, 517 (2006)
[arXiv:hep-ph/0508037].

\bibitem{Dolan:1973qd}
L.~Dolan and R.~Jackiw,
Phys.\ Rev.\ D {\bf 9}, 3320 (1974);
S.~Weinberg,
Phys.\ Rev.\ D {\bf 9}, 3357 (1974).

\bibitem{Kibble:1976sj}
T.~W.~B.~Kibble,
J.\ Phys.\ A {\bf 9}, 1387 (1976).

\bibitem{Zurek:1996sj}
W.~H.~Zurek,
Phys.\ Rept.\  {\bf 276}, 177 (1996)
[arXiv:cond-mat/9607135].

\bibitem{Einhorn:1980ym}
M.~B.~Einhorn, D.~L.~Stein and D.~Toussaint,
Phys.\ Rev.\ D {\bf 21}, 3295 (1980).

\bibitem{Dicus:1982ri}
D.~A.~Dicus, D.~N.~Page and V.~L.~Teplitz,
Phys.\ Rev.\ D {\bf 26} (1982) 1306.

\bibitem{Kolb:1990vq}
E.~W.~Kolb and M.~S.~Turner,
Redwood City, USA: Addison-Wesley (1990) 547 p. (Frontiers in physics, 69).

\bibitem{Lazarides:1980tf}
G.~Lazarides, Q.~Shafi and T.~F.~Walsh,
Phys.\ Lett.\ B {\bf 100}, 21 (1981).

\bibitem{Turner:1982ag}
M.~S.~Turner, E.~N.~Parker and T.~J.~Bogdan,
Phys.\ Rev.\ D {\bf 26}, 1296 (1982).

\bibitem{Moroi:1994rs}
T.~Moroi, M.~Yamaguchi and T.~Yanagida,
Phys.\ Lett.\ B {\bf 342}, 105 (1995)
[arXiv:hep-ph/9409367]; M.~Kawasaki, T.~Moroi and T.~Yanagida,
Phys.\ Lett.\ B {\bf 370}, 52 (1996)
[arXiv:hep-ph/9509399].

\bibitem{Rajagopal:1990yx}
K.~Rajagopal, M.~S.~Turner and F.~Wilczek,
Nucl.\ Phys.\ B {\bf 358}, 447 (1991);  A.~Brandenburg, L.~Covi, K.~Hamaguchi, L.~Roszkowski and F.~D.~Steffen,
Phys.\ Lett.\ B {\bf 617}, 99 (2005)
[arXiv:hep-ph/0501287] and refrences therein.

\bibitem{Fukugita:1986hr}
M.~Fukugita and T.~Yanagida,
Phys.\ Lett.\ B {\bf 174}, 45 (1986); For non-thermal leptogenesis see G.~Lazarides and Q.~Shafi,
Phys.\ Lett.\ B {\bf 258}, 305 (1991).

\bibitem{Panagiotakopoulos:1987he}
C.~Panagiotakopoulos and Q.~Shafi,
Phys.\ Lett.\ B {\bf 197}, 519 (1987).

\bibitem{Dimopoulos:1987rk}
S.~Dimopoulos and L.~J.~Hall,
Phys.\ Lett.\ B {\bf 196}, 135 (1987); L.~M.~Krauss and M.~Trodden,
Phys.\ Rev.\ Lett.\  {\bf 83}, 1502 (1999)
[arXiv:hep-ph/9902420]; S.~Davidson, M.~Losada and A.~Riotto,
Phys.\ Rev.\ Lett.\  {\bf 84}, 4284 (2000)
[arXiv:hep-ph/0001301]; H.~D.~Kim, J.~E.~Kim and T.~Morozumi,
Phys.\ Lett.\ B {\bf 616}, 108 (2005)
[arXiv:hep-ph/0409001]; K.~S.~Babu, R.~N.~Mohapatra and S.~Nasri,
[arXiv:hep-ph/0606144].

\bibitem{Affleck:1984fy}
I.~Affleck and M.~Dine,
Nucl.\ Phys.\ B {\bf 249}, 361 (1985).

\bibitem{Jeong:2004hy}
D.~Jeong, K.~Kadota, W.~I.~Park and E.~D.~Stewart,
JHEP {\bf 0411}, 046 (2004)
[arXiv:hep-ph/0406136].

\bibitem{Charng:2005um}
Y.~Y.~Charng, D.~S.~Lee, C.~N.~Leung and K.~W.~Ng,
Phys.\ Rev.\ D {\bf 72}, 123517 (2005)
[arXiv:hep-ph/0506273].

\bibitem{Kuzmin:1985mm}
V.~A.~Kuzmin, V.~A.~Rubakov and M.~E.~Shaposhnikov,
Phys.\ Lett.\ B {\bf 155}, 36 (1985).

\bibitem{Spergel:2006hy}
D.~N.~Spergel {\it et al.} [WMAP Collaboration],
[arXiv:astro-ph/0603449].

\bibitem{King:1997ia}
S.~F.~King and Q.~Shafi,
Phys.\ Lett.\ B {\bf 422}, 135 (1998)
[arXiv:hep-ph/9711288].

\bibitem{Babu:1997js}
K.~S.~Babu, J.~C.~Pati and F.~Wilczek,
Phys.\ Lett.\ B {\bf 423}, 337 (1998)
[arXiv:hep-ph/9712307].

\bibitem{Antoniadis:1988cm}
I.~Antoniadis and G.~K.~Leontaris,
Phys.\ Lett.\ B {\bf 216}, 333 (1989); I.~Antoniadis, G.~K.~Leontaris and J.~Rizos,
Phys.\ Lett.\ B {\bf 245}, 161 (1990).

\bibitem{Balestra:2006fr}
S.~Balestra {\it et al.}  [SLIM Collaboration],
[arXiv:hep-ex/0601019]; S.~Balestra {\it et al.}  [SLIM Collaboration],
PoS {\bf HEP2005}, 018 (2006)
[arXiv:hep-ex/0602036].

\bibitem{Niessen:2001ci}
P.~Niessen,
``Search for relativistic monopoles with the AMANDA detector,''
%
{\it Prepared for 27th International Cosmic Ray Conference (ICRC 2001), Hamburg, Germany, 7-15 Aug 2001.}

\bibitem{Spiering:2005xv}
C.~Spiering,
Phys.\ Scripta {\bf T121}, 112 (2005)
[arXiv:astro-ph/0503122]; A.~Achterberg {\it et al.}  [IceCube Collaboration],
[arXiv:astro-ph/0604450].


\end{thebibliography}
\end{document}